\documentclass[aps,prb,twocolumn,superscriptaddress,showpacks]{revtex4-2}

\usepackage{comment}
\usepackage[colorlinks, citecolor=red]{hyperref}
\usepackage{hyperref}
\usepackage[utf8]{inputenc}
\usepackage[T1]{fontenc}    %
\usepackage[english]{babel}
\usepackage[pdftex]{graphicx}
\usepackage{amsmath,amssymb,amsthm,bbm, mathtools} %
\usepackage{mathrsfs}
\usepackage{tikz}   %
\usepackage[normalem]{ulem}

\renewcommand{\vec}[1]{\boldsymbol{#1}}

\newcommand{\CR}[1]{{\color{red} #1}}

\begin{document}


\title{Renormalized and iterative formalism of the Andreev levels within large multi-parametric space}

\author{Xian-Peng Zhang}

\affiliation{Department of Physics, Hong Kong University of Science and Technology, Clear Water Bay, Hong Kong, China}

\affiliation{Centre for Quantum Physics, Key Laboratory of Advanced Optoelectronic Quantum Architecture and Measurement (MOE), School of Physics, Beijing Institute of Technology, Beijing, 100081, China}

\begin{abstract}
We attain a renormalized and iterative expression of the Andreev level in a quantum-dot Josephson junction, which is bound to have significant implications due to several significant advantages. The renormalized form of the Andreev level not only allows us to extend beyond the limitations of small tunnel coupling, quantum dot energy, magnetic field, and mean-field Coulomb interaction but also enables the capturing of subgap levels that leak out of the superconducting gap into the continuous spectrum. These leaked subgap levels are highly tunable by gate, phase, and field parameters and play a significant role in the novel phenomena and remarkable properties of the superconductor. Furthermore, the iterative form of the Andreev level provides an intuitive understanding of the spin-split and superconducting proximity effects of the superconducting leads.  We find a singlet-doublet quantum phase transition (QPT) in the ground state due to the intricate competition between the superconducting and spin-split proximity effects, that differs from the typical QPT arising from the competition between the superconducting proximity effect (favoring singlet phase) and the quantum dot Coulomb interaction (favoring doublet phase). This QPT has a diverse phase diagram owing to the spin-split proximity effects which favors the doublet phase akin to the quantum-dot Coulomb interaction but can be also enhanced by the tunneling coupling like the superconducting proximity effect. 
Unlike the typical QPT, where tunnel coupling prefers singlet ground state, this novel QPT enables strong tunnel coupling to suppress the singlet ground state via the spin-split proximity effect, allowing a singlet-doublet-singlet transition with increasing tunnel coupling. Our renormalized and iterative formalism of the Andreev level  is crucial for the electrostatic gate, external flux, and magnetic field modulations of the Andreev qubits.
\end{abstract}

\maketitle

\section{Introduction}
The Andreev qubit, featuring with the remarkable scalability of superconducting circuits and the compact footprint of quantum dots, is currently a subject of particular interest \cite{zazunov2003andreev,janvier2015coherent,bretheau2013supercurrent,bretheau2013exciting,chtchelkatchev2003andreev,tosi2019spin,hays2021coherent,wendin2021coherent}. Depending on the occupation of Andreev levels, the even- and odd-parity Andreev states are responsible for the Andreev level qubit  \cite{zazunov2003andreev,janvier2015coherent,bretheau2013supercurrent,bretheau2013exciting} and the Andreev spin qubit  \cite{chtchelkatchev2003andreev,tosi2019spin,hays2021coherent,wendin2021coherent}, respectively. The occupations of the Andreev levels depend on the competition between the superconducting proximity effect  and quantum-dot Coulomb interaction \cite{valentini2021nontopological,franke2011competition,lee2017scaling}. The former privileges a Bogoliubov-like singlet $\vert S\rangle$ -- a superposition of the empty and doubly occupied states (fermionic even parity) \cite{meng2009self}, while the latter favors a one-by-one electron filling (fermionic odd parity), that is, doublet $\vert D\rangle$ \cite{vecino2003josephson,deacon2010tunneling,lim2015shiba}, whose degeneracy can be lifted by applying a Zeeman magnetic field. Thus, with the lower-energy doublet crossing the lower-energy singlet at a critical 
magnetic field, the ground state can evolve a  singlet-doublet quantum phase transition (QPT) in  superconductors coupled
to various types of semiconducting quantum dots~\cite{higginbotham2015parity,jellinggaard2016tuning,lee2014spin,dvir2019zeeman}. Alternatively, the competition between the Kondo correlation and superconductivity has been known to drive a singlet-doublet QPT in a quantum-dot Josephson junctions~\cite{lee2022proposal}. Here, we exploit the ample additional tuning possibilities afforded by the spin-split proximity effect of superconducting lead, which favors the doublet phase akin to the quantum-dot Coulomb interaction \emph{but} can be enhanced by the tunneling coupling like the superconducting proximity effect. We highlight the singlet-doublet QPT due to the intricate competition between superconducting and spin-split proximity effects of superconducting leads in the absence of the quantum-dot Coulomb interaction. This QPT differs from the typical QPT due to the competition between the superconducting proximity effect and quantum-dot Coulomb interaction and should be paid attention in various types of superconducting and magnetized quantum dot~\cite{liu2019semiconductor,vaitiekenas2021zero,kurtossy2021andreev}.

\begin{figure}[t]
\begin{center}
\includegraphics[width=0.98\linewidth]{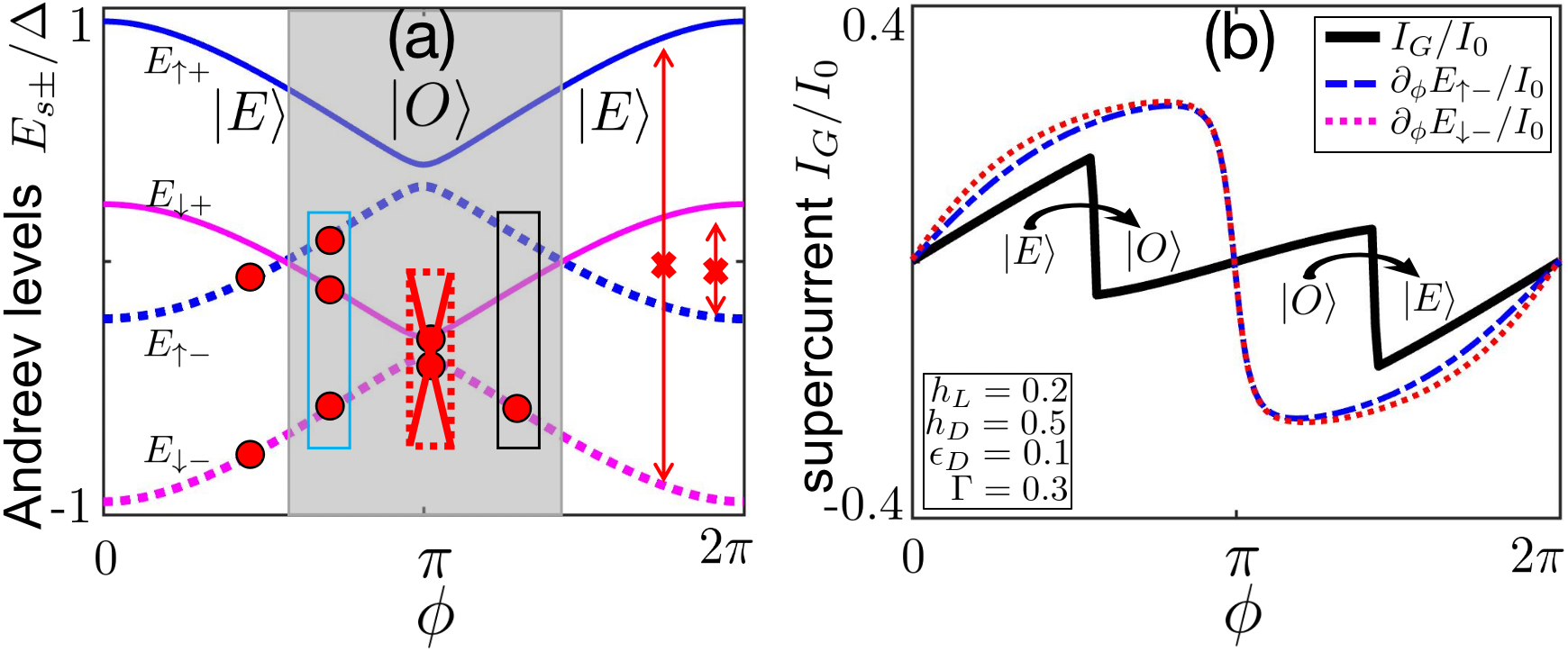}
\end{center}
\caption{(a) The Andreev levels $E_{s\pm}$ as a function of $\phi$.   (b) The ground-state supercurrent $I_G$ as a function of $\phi$. The blue and magenta lines correspond to the subgap supercurrent $\partial_{\phi} E_{\uparrow-}/I_0$ and $\partial_{\phi} E_{\downarrow-}/I_0$, respectively. The sharp drop of total supercurrent [black line in \ref{QPT}(c)] happens when the a pair of Andreev levels dual to each other cross zero [\ref{QPT}(b)] and reveals the fermionic parity change in the ground state.}
\label{SC}
\end{figure}

Microscopic theories of the singlet-doublet QPT rely on the superconducting Anderson model~\cite{rozhkov1999josephson}. However, studying the evolution of the Andreev level within a realistic multi-parametric space often necessitates  prohibitively expensive numerical methods for instance the numerical renormalization group~\cite{bulla2008numerical} and quantum Monte Carlo techniques~\cite{siano2004josephson}.
To tackle this challenge, controlled analytic approximations  have been developed, including mean-field approaches~\cite{yeyati1997resonant,martin2011josephson,yoshioka2000numerical}, functional
renormalization group techniques~\cite{karrasch2008josephson,wentzell2016magnetoelectric}, and perturbation expansions~\cite{vecino2003josephson,meng2009self} within small tunnel coupling, quantum dot energy, magnetic field, and Coulomb interaction limits, where superconductors can be integrated out to generate an effective pair potential on quantum dot~\cite{meng2009self,kurilovich2021microwave,rozhkov1999josephson,bauer2007spectral,meden2019anderson,fatemi2021microwave}. As a result, the integration of superconductor degree of freedom overlooks the entanglement and hybridization between the quantum dot and the superconductors. Additionally, as the tunnel coupling, quantum dot energy, magnetic field, and Coulomb interaction increase, the outer subgap levels leak out of the superconducting gap into the continuous part
of the superconducting spectrum~\cite{Zalom2023rigorous,zalom2022subgap,pavevsivc2023quantum}. Notably, these leaked subgap levels are highly gate-, phase-, and field-tunable (see Appendix~\ref{detivationofALa}) and  thus play a crucial role in the novel phenomena and remarkable properties of the superconductor, for example, the supercurrent arising from the phase-tunable leaked subgap levels. Therefore, an analytical expression of Andreev levels within larger multi-parametric
space is highly desirable and proves valuable for dominating the various  properties of the low-energy superconducting condensate.

By directly diagonalizing an infinite system, we derive a renormalized and iterative expression for the Andreev level in the presence of Zeeman magnetic fields and mean-field Coulomb interaction. Notably, the renormalized formalism of the Andreev level not only overcomes the limitations of small tunnel coupling, quantum dot energy, magnetic field, and Coulomb interaction, but also enables the capture of gate-, flux-, and field-tunable subgap levels that leak out of the superconducting gap into the continuous spectrum. Moreover, the iterative formalism of the Andreev level provides an intuitive understanding of the spin-split and superconducting proximity effects of the superconducting leads. We underline the singlet-doublet QPT in the ground state due to the intricate competition between the superconducting and spin-split proximity effects. This QPT has a diverse phase diagram owing to the spin-split proximity effects, which favors doublet phase akin to the quantum-dot Coulomb interaction but can be also enhanced by the tunneling coupling like the superconducting proximity effect (favoring singlet phase). 
Unlike the typical QPT, where tunnel coupling prefers singlet ground state, this novel QPT enables strong tunnel coupling to suppress the singlet ground state via the spin-split proximity effect, allowing a singlet-doublet-singlet transition with increasing tunnel coupling.

The paper is organized as follows. In Sec.~\ref{modelandtheory}, we present our model and theory.  Section~\ref{resultsanddiscussions} presents our results and discussions, including superconducting and spin-split proximity effects (Sec.~\ref{Resultcompetition}), QPT in the absence of Coulomb interaction (Sec.~\ref{Resultabsence}), as well as  QPT in the presence of Coulomb interaction (Sec.~\ref{Resultpresence}). Our paper ends with conclusion and acknowledgement in Sec.~\ref{conclusion} and Sec.~\ref{acknowledgement}.  Finally, Appendix~\ref{detivationofALa}, and  Appendix~\ref{YushibaRusinov} present the microscopic derivations of Andreev levels in the absence and presence of the quantum-dot Coulomb interaction, respectively.

\section{Model and theory} \label{modelandtheory}
We study the hybrid quantum dot and superconducting lead system containing Zeeman magnetic fields.  The dynamics of the hybrid system can be captured by the following Hamiltonian~\cite{van2017magnetic,jellinggaard2016tuning,gediminas2015yu}
\begin{align} \label{qfdvkfvk}
    H=H_L+H_D+H_T.
\end{align}
The quantum dot Hamiltonian, in the presence of  on-site Coulomb interaction $U$, is given by 
\begin{align} \label{qfdkvmdk}
    H_D=\sum_{s=\uparrow,\downarrow}(\epsilon_D+sh_D)n_{ s}^{}+Un_{\uparrow}n_{\downarrow}.
\end{align}
Here, $n_s=d^{\dagger}_sd^{}_s$ is number operator with spin $s$, and $d_{ s}^{\dagger}$ is the creation operator of the electron with spin $s$ and energy $\epsilon_D+sh_D$. $\epsilon_D$ and $h_D$ are the energy level and Zeeman energy in the quantum dot, respectively.
The leads are conventional singlet $s$-wave superconductors and represented by the  Bardeen-Cooper-Schrieffer Hamiltonian 
\begin{align} \label{qfdvlkkld}
    H_{L}&=\sum_{jn\vec{k}s}(\epsilon_{jn\vec{k}}+sh_L)c^{\dagger}_{jn\vec{k}s}c^{}_{jn\vec{k}s}\\
    &+\sum_{j\vec{k}}\left(\Delta_j c^{\dagger}_{jn\vec{k}\uparrow}c^{\dagger}_{jn-\vec{k}\downarrow}+h.c\right), \notag 
\end{align}
where $c_{jn\vec{k}s}$ is the annihilation operator of superconductor $j=1,2$ with band $n$, spin $s$, wave vector $\vec{k}$, and energy $\epsilon_{jn\vec{k}s}$, where $\epsilon_{jn\vec{k}}$ and $h_L$ are the energy spectrum and Zeeman energy in superconductor $j$, respectively. The pair potentials of two superconductors, $\Delta e^{i\phi_j}$ have a phase difference $\phi=\phi_1-\phi_2$. The leads and the dots are tunnel-coupled as follows 
\begin{align} \label{qdvldl}
    H_T=\sum_{jn\vec{k}s}\left(t_jc^{\dagger}_{jn\vec{k}s}d^{}_{s}+h.c.\right).
\end{align}
Assumed to be real and spin-, band-, and momentum-independent, the tunnel coupling amplitude $t_j$ is generated by the overlap between the wave functions in the nanowire and quantum dot \cite{probst2016signatures}.

Strictly speaking, for a large enough field, the pair potential has to be determined self-consistently, and striking phenomena, such as the inhomogeneous superconducting phase  \cite{larkin2005theory,fulde1964superconductivity},  might appear. The situation becomes simpler when superconductivity and magnetism arise from the proximity effects without any self-consistent calculation~\cite{zhang2020phase}. Thus, we study the situation in which a nanowire is in contact with superconductors and ferromagnetic insulators~\cite{liu2019semiconductor,vaitiekenas2021zero,kurtossy2021andreev}, where ferromagnetic proximity effect can extend into the nanowire proximitized by superconductors for a superconducting coherence length and thin enough nanowire becomes uniformly magnetized \cite{zhang2020phase,bergeret2004induced}. Hereafter, we still call the nanowire proximitized by superconductors as superconducting leads for briefness. Besides, we consider the collinear but distinguishable Zeeman magnetic fields in quantum dot ($h_D$) and superconducting leads ($h_L$), and the global chemical potential is set to be zero for briefness. We describe the tunneling between the quantum dot and superconductor $j$ with $\Gamma_j=\pi\nu_Ft^2_j$, where $\nu_F$ is the density of state of the superconductors.

Treating Coulomb interaction in a mean-field way, the {\it quadratic} Hamiltonian \eqref{qfdvkfvk} of the hybrid system  is exactly solvable~\cite{de2018superconductivity,zhang2024fermi} 
\begin{align} \label{fdbldfl}
   H&=\frac{1}{2}\sum^{N}_{l=1}\sum_{s=\uparrow/\downarrow}\sum_{\eta=+/-} E^{}_{ls\eta}\gamma_{ls\eta}^{\dagger} \gamma_{ls\eta }^{}+\mathcal{E}.
\end{align} 
Here, $s=\uparrow/\downarrow$ is for spin-up and -down in the absence of the spin-flip from spin-orbit coupling and spin-dependent tunneling. The additional index $\eta=+/-$ labels the high/low energy levels of each spin species which satisfy  $E^{}_{ls+}>E^{}_{ls-}$ and $E_{l\uparrow\eta}>E_{l\downarrow\eta}$. $\mathcal{E}$ is  constant energy. The quasiparticle operator $\gamma_{ls\eta }^{}$ with energy $E_{ls\eta}$ is a unit vector in Nambu space of the hybrid 
system 
\begin{align} \label{tfvvldl} 
     \gamma_{ls\eta}= \sum_{k}\left[(u_{s\eta})^{ }_{lk}c^{}_{ks}+(v_{s\eta})^{}_{lk}(-sc^{\dagger}_{k-s})\right].
\end{align}
Here, we denote $c_{ks}$ such that $c_{1s}=d_s$ and $c_{ks}=c_{j\vec{k}s}$ for all $k>1$. $(u_{s\eta})^{ }_{lk}$ and $(v_{s\eta})^{}_{lk}$ describe the electron  and hole  distributions of quasiparticles $(\gamma_{ls\eta})$, respectively. 

Note that $\{\gamma^{}_{ls\eta}\}$ for all $l$, $s$, and $\eta$  is an overcomplete basis set including two orthonormal basis sets dual to each other  and the quasiparticle states  satisfy conjugate relation $\gamma^{\dagger}_{ls\eta}= \gamma^{}_{l-s-\eta}$ and particle-hole symmetry $E_{ls\eta}=-E_{l-s-\eta}$. Next, we divide the overcomplete basis set into two orthonormal basis sets dual to each other --  $\{\gamma^{}_{ls+}\}$ for all $l$ and $s$ as well as $\{\gamma^{}_{ls-}\}$ for all $l$ and $s$. Defined as $\gamma_{ls\eta}\vert V\rangle_{\eta}=0$ for  all $l$ and $s$, the effective vacuum states can be rewritten to the Bogoliubov-like singlet form
\begin{equation} \label{trufvdfmmvl}
    \vert V\rangle_{\eta}=\frac{1}{N^{1/2}_{\eta}}\prod^{N}_{k=1}\left(1+\mathcal{A}^{\eta}_{k}a^{\dagger}_{k\uparrow\eta}a^{\dagger}_{k\downarrow\eta}\right)\vert 0\rangle,
\end{equation}
where $N_{\eta}=\prod_{k}\left(1+\vert\mathcal{A}^{\eta}_{kk}\vert^2\right)$. The expressions of Andreev coefficients $\mathcal{A}^{\eta}_{k}$ and superconducting spin clouds $a^{}_{ks\eta}$ are given in Ref. \cite{zhang2024fermi}.  The ground state is the filled Fermi sea of all the negative quasiparticles in the $\eta=+$ orthonormal basis set starting from the effective vacuum state $\vert V \rangle_+$. The same applies to the $\eta=-$ orthonormal basis set. Therefore, we reach
\begin{align} \label{yfvklal}
    \vert G \rangle=\left(\prod_{E_{ls+}<0}\gamma^{\dagger}_{ls+}\right)\vert V\rangle_+=\left(\prod_{E_{ls-}<0}\gamma^{\dagger}_{ls-}\right)\vert V\rangle_-,
\end{align}
whose ground-state energy is given by 
\begin{align} \label{groundstateenergy}
    \mathcal{E}_G=\mathcal{E}_{+}+\sum_{E_{ls+}<0}E_{ls+}=\mathcal{E}_{-}+\sum_{E_{ls-}<0}E_{ls-}.
\end{align}
Here, $\mathcal{E}_{\eta}=\mathcal{E}+\sum_{\eta'\neq\eta}\frac{1}{2}E^{}_{ls\eta'}$ is  the effective vacuum state energy of the corresponding orthonormal basis set.
Note that superconducting spin clouds $a^{}_{ks\eta}$ include both quantum dot and superconducting degrees of freedom~\cite{zhang2024fermi}, and therefore our ground state \eqref{yfvklal} and effective vacuum state \eqref{trufvdfmmvl} capture the entanglement and hybridization between the quantum dot and the superconductors. Therefore, we demonstrate the singlet-doublet QPT from wave-function perspective.

\section{Results and discussions} \label{resultsanddiscussions}

\subsection{Superconducting and spin-split proximity effects} \label{Resultcompetition}
Hereafter, we mainly focus on Andreev space -- the low-energy subgap space of the hybrid system [$l=1$ of Eq. \eqref{fdbldfl}], described by the Andreev levels, $E_{s\eta}=E_{1s\eta}$, which can be obtained from the implicit equation (see detailed derivations in Appendix~\ref{detivationofALa})
\begin{equation} \label{fvnfdl}
     E_{s\eta}=s\frac{h_D+\tilde{\Gamma}_s(E_{s\eta})h_L}{1+\tilde{\Gamma}_s(E_{s\eta})}+\eta \frac{\sqrt{\epsilon_D^2+\Delta^2\cos^2(\frac{\phi}{2})\tilde{\Gamma}^2_s(E_{s\eta})}}{1+\tilde{\Gamma}_s(E_{s\eta})},
\end{equation}
iteratively, with $\tilde{\Gamma}_s(E_{s\eta})=2\Gamma/\sqrt{\Delta^2-(E_{s\eta}-sh_L)^2}$. Here, we have assumed $\Gamma_j=\Gamma$ for briefness.  $\tilde{\Gamma}_s(E_{s\eta})h_L$ and $\tilde{\Gamma}_s(E_{s\eta})\Delta\cos(\frac{\phi}{2})$ correspond to the spin-split and superconducting proximity effects, respectively. The detailed gate, phase, and field dependence of the Andreev levels \eqref{fvnfdl} is plotted in Appendix~\ref{detivationofALa}, where we show a perfect matching in analytical and numerical Andreev levels and simplified expressions is given in several parameter regimes. Intuitively showing the spin-split and superconducting proximity effects, our iterative expression of the Andreev level \eqref{fvnfdl} simplifies the calculation of the critical conditions of the QPT where the implicit equation become explicit after setting $E_{s\eta}=0$ [see Eq. \eqref{yrfndkvmkf}].  Importantly,  the Andreev levels \eqref{fvnfdl} are renormalized by  $1/[1+\tilde{\Gamma}_s(E_{s\eta})]$, making it possible to go beyond small tunnel coupling, quantum dot energy and magnetic field limits. Noting that $\tilde{\Gamma}_s(E_{s\eta})$ becomes large when Andreev level approaches gap edges, the renormalization effect always forces the Andreev levels $E_{s+}$ and $E_{s-}$ inside $(sh_L-\Delta,sh_L+\Delta) $, as shown by Eq.~\eqref{dfvakv} in Appendix~\ref{detivationofALa}. Moreover, the renormalization effect makes it works when Andreev levels leak out of the superconducting gap $[-\Delta+h_L, +\Delta-h_L]$ into the continuous part
of the superconducting spectrum $(-h_L-\Delta,-\Delta+h_L)$ and $(+\Delta-h_L,+h_L+\Delta)$. These leaked Andreev levels are  highly gate-, phase-, and field-tunable (see Fig.~\ref{FIG2} in Appendix~\ref{detivationofALa}), and hence play a significant role in novel phenomena and remarkable properties of the superconductor. Moreover, as an elegant generalization of the conventional subgap level at zero magnetic field $E_{\eta}=\eta\Delta_{\text{A}} \sqrt{1-\tau_{\text{A}} \sin ^{2}(\phi / 2)}$~\cite{janvier2015coherent,bretheau2013supercurrent}, we can find the microscopic expressions of transmission probability $\tau_{\text{A}}=\frac{\Delta^2\tilde{\Gamma}^2(E_{\eta})}{\epsilon_D^2+\Delta^2\tilde{\Gamma}^2(E_{\eta})}$ and induced minigap $\Delta_{\text{A}}=\frac{\sqrt{\epsilon_D^2+\Delta^2\tilde{\Gamma}^2(E_{\eta})}}{1+\tilde{\Gamma}(E_{\eta})}$, clearly showing the interaction between dot energy $\epsilon_D$ and superconducting proximity effect $\Delta\tilde{\Gamma}(E_{\eta})$. Besides, both contain the renormalization factor and hence work in the strong tunnel coupling, quantum dot energy and magnetic field limits.

\begin{figure}[t]
\begin{center}
\includegraphics[width=0.97\linewidth]{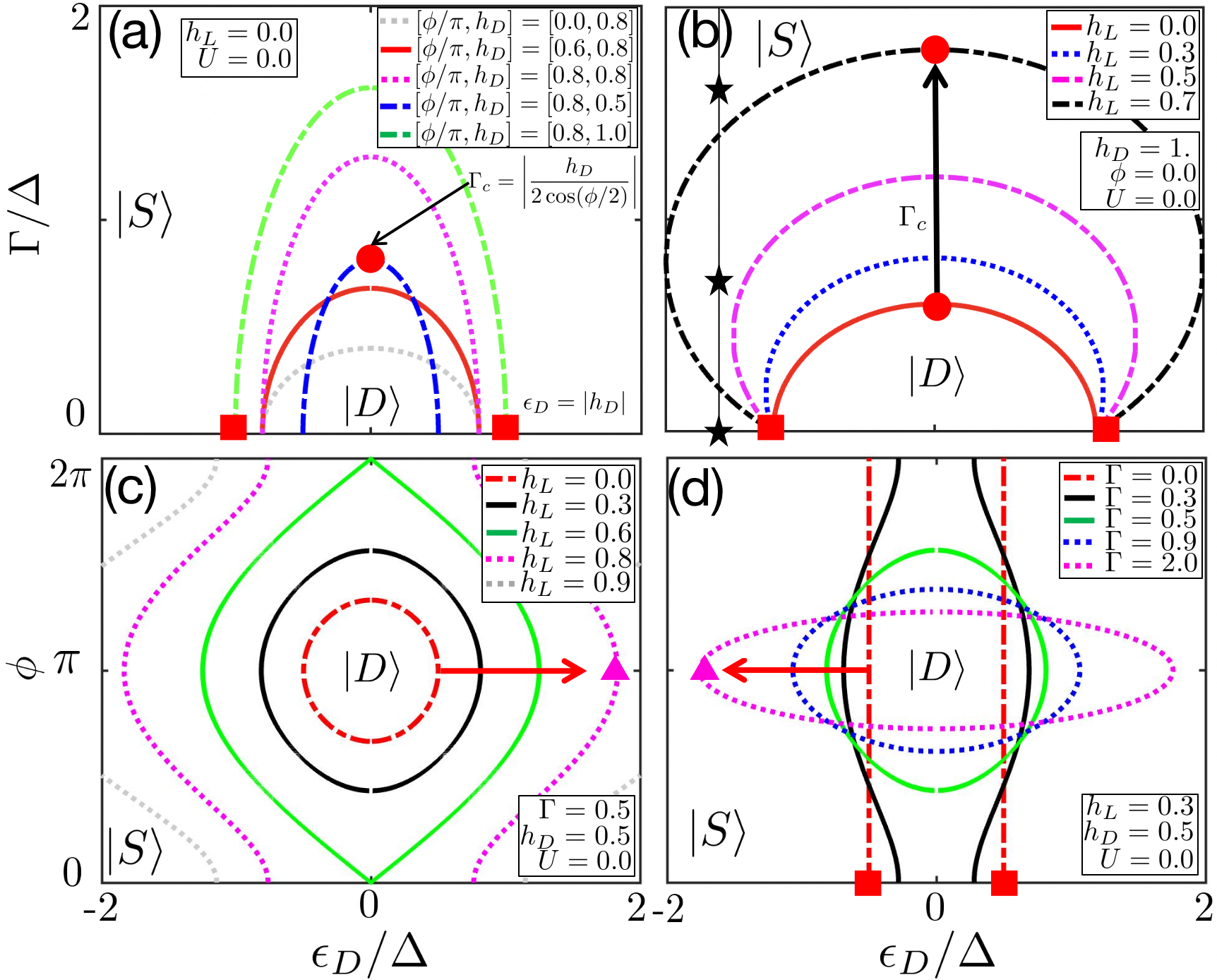}
\end{center}
\caption{The gate-, phase-, and field- tunable singlet-doublet QPT  in the absence of Coulomb interaction. (a-d) The phase diagram as a function of $\epsilon_D$ and (a,b) $\Gamma$ [(c,d)  $\phi$], in the absence of Coulomb interaction [Eq. \eqref{yrfndkvmkf}], where the singlet ($\vert S\rangle $) and doublet ($\vert D\rangle $) sectors are separated by the zero-energy contour $E_{\downarrow+}(\epsilon_D,\Gamma)=0$ [$E_{\downarrow+}(\epsilon_D,\phi)=0$], respectively.
The superconducting [spin-split] proximity effect, parameterized by $\tilde{\Gamma}(0)\Delta\cos(\frac{\phi}{2})$ [$\tilde{\Gamma}(0)h_L$], enhances the singlet [doublet] ground state [panels (a) and (b)]. At $\phi=\pi$, the superconducting proximity effects from two superconductors interfere destructively [Eq. \eqref{yrfndkvmkf}], while the spin-split proximity effect,  increasing with both $h_L$ and $\Gamma$, enhances the doublet ground state [panels (c) and (d)].}
\label{QPT}
\end{figure}

\subsection{QPT in the absence of Coulomb interaction} \label{Resultabsence}
We study the singlet-doublet QPT resulting from the intricate competition between superconducting and spin-split proximity effects. The later promotes the doublet ground state akin to the quantum dot Coulomb repulsion. 
Note that an Andreev bound state of $E_{s\eta}$ contributes to the supercurrent an amount $I_{s\eta}=\frac{2e}{\hbar}\partial_ \phi E_{s\eta}$, where  $e$ is the charge of the electron and $\hbar$ is the reduced Plank constant.  Therefore, the
underlying QPT manifests itself by a sudden change of the ground-state supercurrent $I_{G}=\frac{2e}{\hbar}\partial_{\phi}\mathcal{E}_G$~\cite{zhang2024fermi,zhang2024fabry}. Figure \ref{SC}(b) depicts the ground-state supercurrent $I_G$ as a function of $\phi$. The sharp drop in the ground-state supercurrent occurs when a pair of Andreev levels cross zero \CR{[Fig.~\ref{SC}(a)]}. The difference between solid and dashed curves of Fig.~\ref{SC}(b) unveils the unignorable contributions of continuous spectrum and hence it is necessary to include the continuous and subgap levels in the ground-state energy \eqref{groundstateenergy} to guarantee the uniqueness of the ground-state supercurrent. By following the logical framework based on an overcomplete basis set including both positive and negative orthonormal basis sets~\cite{zhang2024fermi}, we can either add the negative $E_{\downarrow+}$ quasiparticle (cyan box)  or  remove the positive $E_{\downarrow+}$  quasiparticle (black box) which switches fermionic parity [Eq.~\eqref{yfvklal}]  (see another explanation in Ref.~\cite{sakurai1970comments}).  However, we cannot do both (red box) because it violates the Pauli exclusion principle, i.e., $\gamma^{\dagger}_{\downarrow+}\gamma^{}_{\uparrow-}=\gamma^{\dagger}_{\downarrow+}\gamma^{\dagger}_{\downarrow+}=0$. Quantitatively, the fermionic parity of the ground state changes in the presence of negative $E_{ls+}$, as shown in Eq. \eqref{yfvklal}. The first available negative  $E_{ls+}$, if present, is the Andreev level \eqref{fvnfdl}, whose iterative form simplifies the derivation of the critical conditions of the QPT, for example, critical dot energy 
\begin{equation} \label{yrfndkvmkf}
    \epsilon^{c,\pm}_D=\pm \sqrt{[h_D+\tilde{\Gamma}(0)h_L]^2-\Delta^2\tilde{\Gamma}^2(0)\cos^2\left(\frac{\phi}{2}\right)},
\end{equation}
where $\tilde{\Gamma}(0)=2\Gamma/\sqrt{\Delta^2-h_L^2}$. For $h_L=h_D$, we obtain the critical magnetic field $h^{c,\pm}=\frac{\sqrt{\epsilon_D^2+\Delta^2\cos^2(\frac{\phi}{2})\tilde{\Gamma}^2_s(0)}}{1+\tilde{\Gamma}_s(0)}$, whose renormalization effect distinguishes with the critical magnetic field of Ref.~\cite{valentini2021nontopological}. 
Figure~\ref{QPT} plots the phase diagram as a function of $\epsilon_D$ and (a,b) $\Gamma$ [(c,d) $\phi$]. The critical dot energy~\eqref{yrfndkvmkf} reduces to $\epsilon^{c,\pm}_D\simeq \pm\sqrt{h^2_D-4\Gamma^2\cos^2(\phi/2)}$ at zero spin-split proximity effect ($h_L=0$), where increasing $\Gamma$ always enhances the singlet phase and there is not doublet ground state anymore for $\Gamma$ larger than $\Gamma_{c}=\vert h_D/[2\cos(\phi/2)]\vert$ [circle in~\ref{QPT}(a)].  Notably, we obtain the typical semicircle phase diagram \cite{meng2009self,vecino2003josephson,rozhkov1999josephson}, but in our case  Coulomb interaction is not necessary.  With increasing $h_L$, we find larger $\Gamma_{c}$ [black arrow in~\ref{QPT}(b)].  
At $\phi=\pi$, the superconducting proximity effects from two superconductors interfere destructively [Eq. \eqref{yrfndkvmkf}], while the spin-split proximity effect, surviving at $\phi=\pi$, causes renormalization of the dot field. Then,  Eq. \eqref{yrfndkvmkf} reduces to $\epsilon^{c,\pm}_D=\pm \vert h_D+\tilde{\Gamma}(0)h_L \vert$ (triangles), which increases with  $h_L$  and $\Gamma$  [red arrows in~\ref{QPT}(c) and \ref{QPT}(d)]. The latter implies we can even {\it enhance}  the doublet ground state by increasing $\Gamma$ for $\phi$ around $\pi$ [\ref{QPT}(d)] which is opposite to the typical singlet-doublet QPT based on the quantum dot Coulomb interaction where strong tunnel coupling favors the singlet ground state. Moreover, this enhancement even happens at maximum superconducting proximity effect ($\phi=0$) owing to the strong spin-split proximity effect [black and magenta lines in \ref{QPT}(b)]. Our novel phase diagram, going beyond the typical phase diagram, allows a $\vert S \rangle-\vert D \rangle-\vert S \rangle$ transition with increasing tunnel coupling [black stars in \ref{QPT}(b)]. Moreover, our iterative formalism allows for precise control of the ground-state phase. For example, we achieve pure doublet ground state for all $\phi$ [\ref{QPT}(c) and \ref{QPT}(d)],  which is crucial for the flux-tunable Andreev spin qubit.

\subsection{QPT in the presence of Coulomb interaction} \label{Resultpresence}

Though we obtain the semicircle phase diagram {\it without} Coulomb interaction [Fig. \ref{QPT}(a)], an unavoidable question is how to distinguish it with the typical semicircle phase diagram from Coulomb interaction in realistic experiments. Next, we include a quantum dot Coulomb interaction $U$ and study the Yu-Shiba-Rusinov case within the  Hartree-Fock-Bogoliubov approximation \cite{valentini2021nontopological,lee2014spin,vecino2003josephson}. The expressions of Andreev levels are given by the iterative form (see detailed derivations in Appendix~\ref{YushibaRusinov})
\begin{align} \label{tfkvldblf}
    E_{s\eta}&=s\frac{h^r_D+\tilde{\Gamma}_s(E_{s\eta})h_L}{1+\tilde{\Gamma}_s(E_{s\eta})}\\
    &+\eta \frac{\sqrt{\left(\epsilon^r_D\right)^2+\left[\Delta^r_D-\Delta\cos(\frac{\phi}{2})\tilde{\Gamma}_s(E_{s\eta})\right]^2}}{1+\tilde{\Gamma}_s(E_{s\eta})}.\notag 
\end{align}
The Coulomb interaction leads to the corrections in dot energy $\epsilon^r_D=\epsilon_D+\frac{U}{2}+\frac{U}{2}\sum_{l\eta}f\left(E_{ls\eta}\right)\frac{\partial}{\partial \epsilon^r_D}E_{ls\eta}$, dot field $h^r_D=h_D+s\frac{U}{2}-\frac{U}{2}\sum_{l\eta}f\left(E_{ls\eta}\right)\frac{\partial}{\partial h^r_D}E_{ls\eta}$, and dot pair potential $\Delta^r_D=-\frac{U}{2}\sum_{l\eta}f\left(E_{ls\eta}\right)\frac{\partial}{\partial \Delta^r_D}E_{ls\eta}$. Here, $f(E)=1/(1+e^{E/k_BT})$ is the Fermi–Dirac distribution at temperature $T$, where $k_B$ is the Boltzmann constant.

\begin{figure}[t]
\begin{center}
\includegraphics[width=1\linewidth]{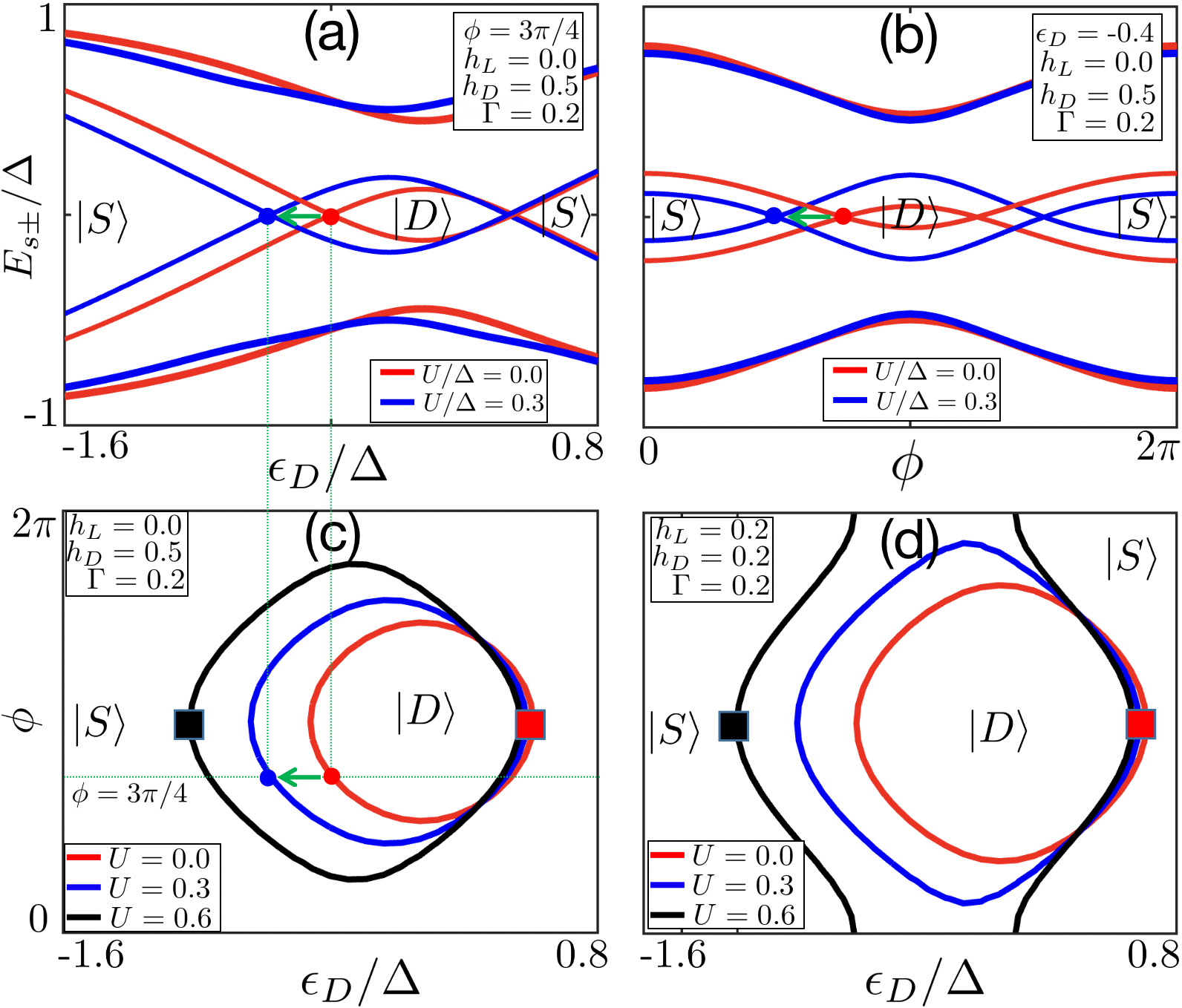}
\end{center}
\caption{The gate-, phase-, and field- tunable singlet-doublet QPT in the presence of Coulomb interaction. (a,b) Andreev levels \eqref{fkvldblf} as a function  (a) $\epsilon_D$ and (b) $\phi$ for different Coulomb interactions. Here, we set the temperature to zero. (e,f) The phase diagram as a function of $\epsilon_D$ and $\phi$ for (e) $h_L/\Delta=0.0$ and (f) $h_L/\Delta=0.2$, respectively, in the presence of Coulomb interaction [Eq. \eqref{tytrgjtor}].  The Coulomb interaction, preferring doublet occupancy, enhances the region of the doublet ground state. Furthermore, we find that the positive critical dot energy \eqref{tytrgjtor} is almost independent of $U$ at $\phi=\pi$ and $T\rightarrow 0$K (red boxes), while the negative critical dot energy \eqref{tytrgjtor} is shifted by Coulomb interaction (black boxes).  }
\label{FIGCI}
\end{figure}

Figures~\ref{FIGCI} (a) and (b), respectively, plot the Andreev levels as a function of $\epsilon_D$ and $\phi$. We see a clear zero-energy shift (green arrows) due to the Coulomb interaction preferring doublet ground state. This shift enhances  the region of the doublet ground state as shown in Fig.~\ref{FIGCI} (c), which is further enhanced by the spin-split proximity effect ($h_L\neq 0$) [Fig.~\ref{FIGCI} (d)]. Moreover, we find the Coulomb enhancement of the doublet ground state relies on the phase difference. Take $\phi=\pi$ as an example, where the superconducting proximity effects from two superconductors interfere destructively. We attain critical dot energy in small tunnel coupling, quantum dot energy, magnetic field, and Coulomb interaction limits (see detailed derivations in Appendix~\ref{YushibaRusinov})
\begin{equation} \label{tytrgjtor}
   \epsilon^{c,\pm}_D
\simeq\pm\left[h_D+\tilde{\Gamma}(0)h_L -Uf(E_{\uparrow+})\frac{\partial}{\partial h^r_D}E_{\uparrow+}\right]-\delta^{-}_{\pm}U.
\end{equation}
Equation \eqref{tytrgjtor} at zero temperature reduces to $\epsilon^{c,\pm}_D\simeq \pm(h_D+\tilde{\Gamma}(0)h_L)-\delta^{-}_{\pm}U$, where the positive critical dot energy $\epsilon^{c,+}_D\simeq h_D+\tilde{\Gamma}(0)h_L$ is almost independent of Coulomb interaction (red boxes), while the negative critical dot energy $\epsilon^{c,-}_D\simeq -h_D-\tilde{\Gamma}(0)h_L-U$ is shifted by Coulomb interaction (black boxes). Therefore, this QPT features with asymmetric behavior with dot energy $\epsilon_D$, i.e., $E_{s\eta}(-\epsilon_D)\neq E_{s\eta}(-\epsilon_D)$, enabling distinguish it from the symmetric QPT due to the interplay of superconducting and spin-split proximity effects (Fig. \ref{QPT}). Though this  asymmetry can be artificially removed by  redefining dot energy $\tilde{\epsilon}_D=\epsilon_D+U/2$, i.e., $E_{s\eta}(-\tilde{\epsilon}_D)= E_{s\eta}(-\tilde{\epsilon}_D)$, we can still use this effect in the experiments where the exact value of $\epsilon_D$ is known and controllable. For example, the asymmetry of the QPT with $\epsilon_D$, in principle, offers an  estimation of the quantum dot Coulomb interaction.  The gate \cite{pillet2010andreev,jellinggaard2016tuning,lee2017scaling}-, field \cite{lee2014spin,whiticar2021zeeman}-, phase \cite{fatemi2021microwave,chang2013tunneling}-tunable QPT has been observed in recent experiments. Though the experimental data are well explained by the Coulomb blockade, our theory might provide a more exhaustive illustration.

\section{Conclusion} \label{conclusion}
We highlight a renormalized and iterative expression of the Andreev level in a quantum-dot Josephson junction, which is bound to have significant implications due to several significant advantages as follows. 
Firstly, the iterative form of the Andreev level provides an intuitive understanding of the spin-split and superconducting proximity effects of the superconducting leads, while also effectively incorporating corrections for the effective quantum dot energy, magnetic field, and pair potential resulting from the quantum dot Coulomb interaction. Importantly, our iterative form of the Andreev level simplifies the analytical calculation of the critical conditions of the singlet-doublet QPT. Secondly, the renormalized form of the Andreev level not only allows us to extend beyond the limits of small tunnel coupling, quantum dot energy level, magnetic field, and Coulomb interaction but also enables the capture of subgap resonances that leak out of the superconducting gap into the continuous spectrum. These leaked subgap resonances are highly tunable by gate, phase, and field parameters and therefore play a significant role in the novel phenomena and remarkable properties of the superconductor. Last but not least, we derive the microscopic expressions for the transmission probability and the induced gap, which clearly shows the interaction between dot energy and superconducting proximity effect and works in the strong tunnel coupling, dot energy, and dot field limits.

\section{Acknowledgement} \label{acknowledgement}
We sincerely thank Yugui Yao, Jose Carlos Egues, Junwei Liu,  Gao Min Tang, Chuan Chang Zeng and Chun Yu Wan for helpful and sightful discussions. This work is supported by the National Natural Science Foundation of China (Grant Nos. 12234003, 12061131002), the National Key R\&D Program of China (Grant No. 2020YFA0308800).

\appendix

\section{Derivation of Andreev levels Eq. (5)} 
\label{detivationofALa} 

In this section, we diagonalize the hybrid quantum dot-superconducting lead system to obtain the compact expressions for the Andreev level energies given in Eq. \eqref{fvnfdl} in the main text. Moreover, we show in detail the gate, phase, and field dependence of the Andreev levels.

We rewrite the total Hamiltonian \eqref{qfdvkfvk} in overcomplete basis -- the Nambu space of the hybrid quantum dot-superconductor system
\begin{align} \label{nambu}
    \Psi=
    \begin{bmatrix}
     d^{}_{\uparrow}\\ 
     -d^{\dagger}_{\downarrow}\\
     d^{}_{\downarrow}\\ 
     +d^{\dagger}_{\uparrow}
    \end{bmatrix}\bigoplus_{\vec{k}jn}
    \begin{bmatrix}
      c^{}_{jn\vec{k}\uparrow}\\
      -c^{\dagger}_{jn-\vec{k}\downarrow}\\
      c^{}_{jn\vec{k}\downarrow}\\
      +c^{\dagger}_{jn-\vec{k}\uparrow}
    \end{bmatrix},
\end{align}
where $\bigoplus_{\vec{k}jn} X_{\vec{k}jn}$ concatenates $X_{\vec{k}jn}$ vertically. Thus, the total Hamiltonian \eqref{qfdvkfvk} can be rewritten into  the Bogoliubov–de Gennes (BdG) Hamiltonian in quadratic form
\begin{align} \label{fdvdjvkds}
    H= \frac{1}{2} \Psi^{\dagger}  \mathcal{H}_{\text{BdG}} \Psi+\mathcal{E},
\end{align}
with 
\begin{align} \label{fdhvdfvjdo}
    \mathcal{H}_{\text{BdG}}=\left[\begin{array}{cccc} 
   \mathcal{H}_{\uparrow}  & 0 \\
    0 & \mathcal{H}_{\downarrow}
    \end{array}\right].
\end{align}
The prefactor $1/2$ and $\phi$-independent constant energy $\mathcal{E}=\epsilon_D+\sum_{jn\vec{k}}\epsilon_{jn\vec{k}}$ arise from rewriting the total Hamiltonian \eqref{qfdvkfvk} in overcomplete basis  \eqref{nambu} \cite{bernevig2013topological}. 
$\mathcal{H}^s$ is the corresponding Hamiltonian matrix in the Nambu space
\begin{align} \label{mvndefnvak}
    \mathcal{H}_s =
 \left[\begin{array}{cccc} 
   \epsilon_D\tau_z+sh_D  & \mathcal{H}_T^{\dagger} \\
    \mathcal{H}_T^{} & \mathcal{H}^{s}_L
    \end{array}\right].
\end{align}
Here, the dot energy and Zeeman energy correspond to the Pauli matrix $\tau_z$ and identity in the Nambu space, respectively. We consider only a single energy level on the dot, assuming that the level spacing to the next dot level is larger than all relevant energy scales. The superconducting lead Hamiltonian \eqref{qfdvlkkld} in matrix form is given by 
\begin{align} \label{vfnvmk} 
\mathcal{H}^{s}_L=\bigotimes_{\vec{k}jn}\begin{bmatrix}
      +\epsilon_{jn\vec{k}}+sh_L & -\Delta_j \\
      -\Delta^*_j &  -\epsilon_{jn-\vec{k}}+sh_L 
 \end{bmatrix},
\end{align}
where $\bigotimes_{\vec{k}jn} X_{\vec{k}jn}$ concatenates $X_{\vec{k}jn}$ diagonally. The quantum tunneling \eqref{qdvldl} reads 
\begin{align}
    \mathcal{H}_T^{}=t_j  \bigoplus_{\vec{k}jn}\tau_{z},
\end{align}
where $\bigoplus_{\vec{k}jn} X_{\vec{k}jn}$ concatenates $X_{\vec{k}jn}$ vertically.

One can diagonalize the matrix of the BdG Hamiltonian \eqref{fdhvdfvjdo}
\begin{align}
    U^{}_{}\mathcal{H}^{}_{\text{BdG}}U^{\dagger} _{}=\bigotimes_{ls\eta}[E_{ls\eta}],
\end{align}
where the transformation matrix $U^{}_{}$ is diagonal in spin space 
\begin{align}
    U _{}=\left[\begin{array}{cccc} 
   U_{\uparrow}  & 0 \\
    0 & U_{\downarrow}
    \end{array}\right].
\end{align}
$U_{s} $ is a $2N\times 2N$ unitary matrix, where $N$ is equal to half of the dimension of the Hamiltonian \eqref{mvndefnvak} and describes the size of the hybrid system.  
Hence, the BdG Hamiltonian  \eqref{mvndefnvak} can be mathematically rewritten into 
\begin{align} \label{gfgblrgk}
    H= \frac{1}{2}\sum^{N}_{l=1}\sum_{s=\uparrow/\downarrow}\sum_{\eta=\pm} E_{ls\eta}\gamma_{ls\eta}^{\dagger} \gamma_{ls\eta }^{}+\mathcal{E},
\end{align}
In the presence of superconductivity, we are required to work in the Nambu space doubling the Hilbert space, and hence an additional index $\eta=+/-$ labels the high/low energy levels of each spin species  ($E^{}_{ls+}>E^{}_{ls-}$), which satisfy $E_{ls\eta}=-E_{l-s-\eta}$. The larger $l$ corresponds to a higher energy level. Then, the terms with $l=1$ are Andreev levels -- the subgap energy levels of the hybrid quantum dot and superconductor system and we here call all other energy levels ($l>1$)  Bogoliubov levels -- the out-of-gap energy levels of the hybrid system.   The quasiparticle operator $\gamma_{ls\eta }^{}$ with energy $E_{ls\eta}$ is a unit vector in Nambu space of the hybrid 
system 
\begin{align} \label{tfvvldl} 
     \gamma_{ls\eta}= \sum_{k}\left[(u_{s\eta})^{ }_{lk}c^{}_{ks}+(v_{s\eta})^{}_{lk}(-sc^{\dagger}_{k-s})\right].
\end{align}
Here, we define $c_{ks}$ such that $c_{1s}=d_s$ and $c_{ks}=c_{j\vec{k}s}$ for all $k>1$. $(u_{s\eta})^{ }_{lk}$ and $(v_{s\eta})^{}_{lk}$, picked out from the unitary matrix $U^s$, describe the electron  and hole  distributions of quasiparticles $(\gamma_{ls\eta})$, respectively.

In general,
it is hard to analytically calculate the energy levels $E_{ls\eta}$ of the hybrid 
system from the following determinant equation:
\begin{align} \label{fvmldfvdl}
    \det\left[\begin{array}{cccc} 
    \epsilon_D\tau_z+sh_D-E^{s}  & \mathcal{H}_T^{\dagger} \\
    \mathcal{H}_T^{} & \mathcal{H}^{s}_L-E^{s}
    \end{array}\right]=0.
\end{align}
To find the  Andreev (subgap) level, we can use the determinant identity for four matrices $A,B,C$, and $D$ \cite{silvester2000determinants}:
\begin{align} \label{fvnvkdk}
    \det {\begin{pmatrix}A&B\\C&D\end{pmatrix}}=\det(D)\det \left(A-BD^{-1}C\right),
\end{align}
where $D$ is assumed to be invertible. Thus, to use Eq. \eqref{fvnvkdk} to solve our determinant equation \eqref{fvmldfvdl}, $\mathcal{H}^{s}_L-E^{s}$  is required to be invertible which is true for the Andreev levels ($ |E^s -s h_L| <\Delta$). Then, the determinant equation \eqref{fvmldfvdl} reduces to 
\begin{align} \label{qfvakfva}
    \det\left[\epsilon_{D}\tau_z+sh_D-E^s-\Sigma^s(E^s)\right]=0.
\end{align}
The self-energy in Eq. \eqref{qfvakfva} is given by 
\begin{equation} \label{tfadavmdakvm}
\Sigma^s(E^s)=\sum_{\vec{k}jn}t^2_j\tau_z 
\left[\begin{array}{cc}
 \mathcal \epsilon_{jn\vec{k}}+s h_L-E^s & -\Delta_j\\
-\Delta^*_j &-\epsilon_{jn-\vec{k}}+sh_L-E^s
\end{array}\right]^{-1}\tau_z.
\end{equation}
The summation over 
momenta $\vec{k}$ in Eq. \eqref{tfadavmdakvm} can be replaced by integration for the in-gap Andreev levels $E^s$,
\begin{align} \label{fdnvkn}
    \sum_{\vec{k},n}F(\epsilon_{jn\vec{k}})=\sum_{n}\int \frac{d\vec{k}}{\Omega_j}F(\epsilon_{jn\vec{k}})=\sum_{n}\int d\epsilon \nu_{jn}(\epsilon)F(\epsilon),
\end{align}
where $\Omega_j$ is the volume of the superconductor $j$ and $\nu_{jn}(\epsilon)$ is the density of states of superconductor $j$ per spin and energy band $n$. There exist many bands with complex energy spectrum and we here assume  energy-independent total density of state by setting $\nu^j_F=\sum_{n} \nu_{jn}(\epsilon)$ for simplicity.
Then, the self-energy \eqref{tfadavmdakvm} becomes 
\begin{align} \label{fadavmdakvm}
\Sigma^s(E^s)&=\sum_{j}\Gamma_j
\left[\begin{array}{cc}
 \mathcal{G}_{js}(E^s) & e^{i\phi_j}\mathcal{F}_{js}( E^s) \\
e^{-i\phi_j}\mathcal{F}_{js}(E^s) & \mathcal{G}_{js}(E^s)
\end{array}\right],
\end{align}
where $\mathcal{G}_{js}(E^s)=(E^s-sh_L)/\sqrt{\vert\Delta_j\vert^{2}-(E^s-sh_L)^{2}}$ and $\mathcal{F}_{js}(E^s)=\vert\Delta_j\vert/\sqrt{\vert\Delta_j\vert^{2}-(E^s-sh_L)^{2}}$.  We parameterize the tunnel coupling strength by $\Gamma_j =\pi\nu^j_Ft^2_j$. The in-gap requirement of the Andreev levels guarantees the invertible $D$ in determinat identity \eqref{fvnvkdk} and the integrable self-energy \eqref{fadavmdakvm}. Hence, one can derive the exact Andreev levels from Eq. \eqref{qfvakfva} with the integrated self-energy \eqref{fadavmdakvm}. This procedure amounts to integrating out the superconducting degrees of freedom (which can be achieved in various ways).


Next, let us solve the reduced determinant equation \eqref{qfvakfva}.  By substitution of Eq. \eqref{fadavmdakvm}, the reduced determinant equation \eqref{qfvakfva} becomes 
\begin{widetext}
\begin{align} \label{fvkgflgb}
    \det
    \begin{bmatrix}
    +\epsilon_D+sh_D-E^s-\sum_j\Gamma_j \mathcal{G}_{js}(E^s)  & -\sum_j\Gamma_je^{+i\phi_j}\mathcal{F}_{js}(E^s) \\
    -\sum_j\Gamma_je^{-i\phi_j}\mathcal{F}_{js}(E^s) & -\epsilon_D+sh_D-E^s-\sum_j\Gamma_j \mathcal{G}_{js}(E^s)
    \end{bmatrix}
    =0.
\end{align}
\end{widetext}
We can rewrite the above determinant equation into
\begin{align} \label{fvnkfkg}
    &\left[sh_D-E^s-\sum_j\Gamma_j \mathcal{G}_{js}(E^s)\right]^2\\
    &=\epsilon^2_D+\left\vert\sum_j\Gamma_je^{+i\phi_j}\mathcal{F}_{js}(E^s)\right\vert^2 .\notag 
\end{align}
To better understand the superconducting proximity effect, we introduce the following dimensionless parameter
\begin{align} \label{fvdglw}
    \tilde{\Gamma}_s(E^s)=\sum_j\tilde{\Gamma}_{js}(E^s)\equiv\sum_j\frac{\Gamma_j}{\sqrt{\vert\Delta_j\vert^2-(E^s-sh_L)^2}}.
\end{align}
Thus, Eq. \eqref{fvnkfkg} reduces to 
\begin{align} \label{yefvnkfkg}
    &\left\{sh^r_D-E^s\left[1+\tilde{\Gamma}_s(E^s)\right]\right\}^2\\
    &=\epsilon_D^2+\left\vert\sum_j\Gamma_je^{+i\phi_j}\mathcal{F}_{js}(E^s)\right\vert^2,\notag
\end{align}
with
\begin{align}
    h^r_D=h_D+\tilde{\Gamma}_s(E^s)h_L.
\end{align}
The $\mathcal{G}_{s}(E^s)$ function renormalizes both Andreev levels and its spin splitting. Therefore, we have obtained a compact implicit equation from which we can easily obtain the Andreev levels explicitly by iterations
\begin{align} \label{gogbokee}
     E_{s\eta}=s\frac{h_D+\tilde{\Gamma}_s(E_{s\eta})h_L}{1+\tilde{\Gamma}_s(E_{s\eta})}+\eta \frac{\sqrt{\epsilon_D^2+\left\vert\sum_j\Gamma_je^{+i\phi_j}\mathcal{F}_{js}(E^s)\right\vert^2 }}{1+\tilde{\Gamma}_s(E_{s\eta})}.
\end{align}
Considering the complexity of the expressions of Andreev levels \eqref{gogbokee}, we first consider the case of $\vert\Delta_j\vert=\Delta$. Then, Eq. \eqref{gogbokee} reduces to
\begin{align} \label{goyugbokee}
     E_{s\eta}&=s\frac{h_D+\tilde{\Gamma}_s(E_{s\eta})h_L}{1+\tilde{\Gamma}_s(E_{s\eta})}\\
     &+\eta \frac{\sqrt{\epsilon_D^2+\frac{\vert\Gamma_1e^{+i\phi_1}+\Gamma_2e^{+i\phi_2}\vert^2}{(\Gamma_1+\Gamma_2)^2}\tilde{\Gamma}^2_s(E_{s\eta})\Delta^2}}{1+\tilde{\Gamma}_s(E_{s\eta})}.\notag 
\end{align}
The interplay of the superconducting proximity effects from two superconducting leads is described by $\vert\Gamma_1e^{+i\phi_1}+\Gamma_2e^{+i\phi_2}\vert^2/(\Gamma_1+\Gamma_2)^2$, which interferes destructively when $\Gamma_j=\Gamma$ and $\phi_1-\phi_2=\pi$.

\begin{figure}[t]
\begin{center}
\includegraphics[width=0.98\linewidth]{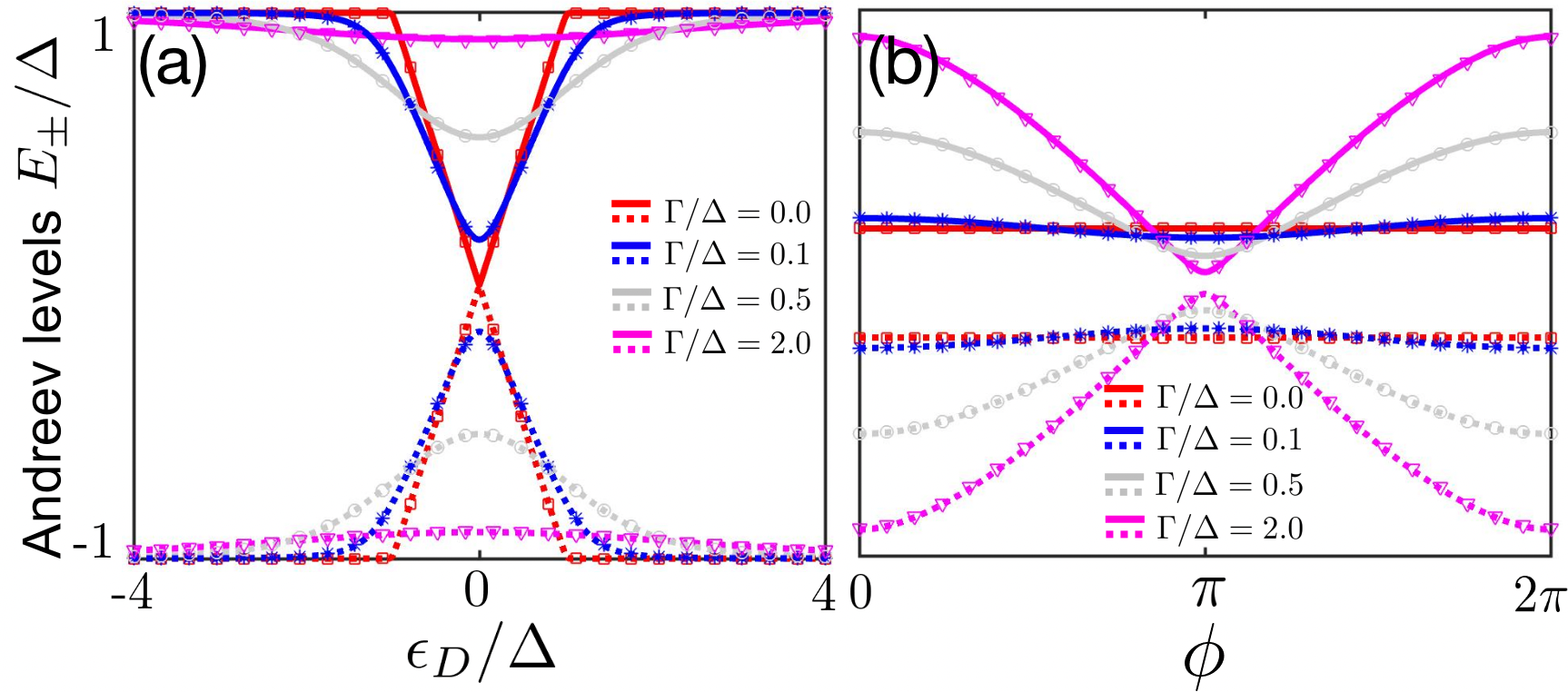}
\end{center}
\caption{Properties of the spin-degenerate Andreev levels. (a,b) Andreev level $E_{\pm}$ as a function of (a) $\epsilon_D$ with $\phi=0$ and (b) $\phi$ with $\epsilon_D/\Delta=0.2$ for different strengths of tunnel coupling $\Gamma$. The lines correspond to the Andreev energies numerically obtained from  Eq. \eqref{mvndefnvak}, which agree well with the analytical Andreev energies \eqref{vfmk} [i.e., the determinant equation \eqref{fvkgflgb}], indicated by the star, square, circle, and triangle marks.}
\label{FIG3}
\end{figure}

Hereafter, we set $\Gamma_j=\Gamma$, $\vert\Delta_j\vert=\Delta$,  $\phi_1=+\frac{\phi}{2}$, and $\phi_2=-\frac{\phi}{2}$ for simplicity. Thus, the Andreev levels \eqref{gogbokee} reduce to 
\begin{align} \label{vfmk}
    E_{s\eta}=s\frac{h_D+\tilde{\Gamma}_s(E_{s\eta})h_L}{1+\tilde{\Gamma}_s(E_{s\eta})}+\eta \frac{\sqrt{\epsilon_D^2+\Delta^2\cos^2(\frac{\phi}{2})\tilde{\Gamma}^2_s(E_{s\eta})}}{1+\tilde{\Gamma}_s(E_{s\eta})},
\end{align}
with
\begin{align}
    \tilde{\Gamma}_s(E^s)=\frac{2\Gamma}{\sqrt{\Delta^2-(E^s-sh_L)^2}}.
\end{align}
For $h_D=0$, we can obtain the spin-split Andreev levels from the spin-split superconductor, while the spin splitting of the quantum dot can be reduced by the  superconducting proximity effect quantified by the renormalization factor  $1/[1+\tilde{\Gamma}_s(E_{s\eta})]$ \cite{futterer2013renormalization}. 
At $\phi=\pm\pi$, the proximity effects from the two superconductors interfere destructively with each other. Thus, the Andreev levels \eqref{vfmk} reduce to 
\begin{align} \label{uvfmk}
    E_{s\eta}=s\frac{h_D+\tilde{\Gamma}_s(E_{s\eta})h_L}{1+\tilde{\Gamma}_s(E_{s\eta})}+\eta \frac{\vert \epsilon_D\vert}{1+\tilde{\Gamma}_s(E_{s\eta})}.
\end{align}
While, in the absence of spin splitting ($h_D=h_L=0$), the Andreev levels \eqref{vfmk} 
become
\begin{align} \label{r}
    E_{\eta}=\eta \frac{\sqrt{\epsilon_D^2+\Delta^2\cos^2(\frac{\phi}{2})\tilde{\Gamma}^2(E_{\eta})}}{1+\tilde{\Gamma}(E_{\eta})},
\end{align}
with
\begin{align}
    \tilde{\Gamma}(E_{\eta})=\frac{2\Gamma}{\sqrt{\Delta^2-E_{\eta}^2}}.
\end{align}
We can find the exact expressions of transmission probability $\tau_{\text{A}}$ and induced gap $\Delta_{\text{A}}$ (not pair potential), defined via $E_{\eta}=\eta\Delta_{\text{A}} \sqrt{1-\tau_{\text{A}} \sin ^{2}(\phi / 2)}$,
\begin{align}
    \tau_{\text{A}}=\frac{\Delta^2\tilde{\Gamma}^2(E_{\text{A}})}{\epsilon_D^2+\Delta^2\tilde{\Gamma}^2(E_{\text{A}})},
\end{align}
\begin{align}
    \Delta_{\text{A}}=\frac{\sqrt{\epsilon_D^2+\Delta^2\tilde{\Gamma}^2(E_{\text{A}})}}{1+\tilde{\Gamma}(E_{\text{A}})}.
\end{align}
These expressions are useful for the quantification of microscopic parameters.   Figure~\ref{FIG3} (a) and (b) plot the zero-magnetic-field Andreev levels $E_{\pm}$, numerically obtained from the diagonalization of Eq. \eqref{mvndefnvak},  as a function of $\epsilon_D$ [$\phi$] for different $\Gamma$, respectively. The numerical Andreev energies agree well with the analytical Andreev energies \eqref{vfmk} indicated by the star, square, circle, and triangle marks [Figs. \ref{FIG3} (a) and (b)].

\begin{figure*}[t]
\begin{center}
\includegraphics[width=0.85\linewidth]{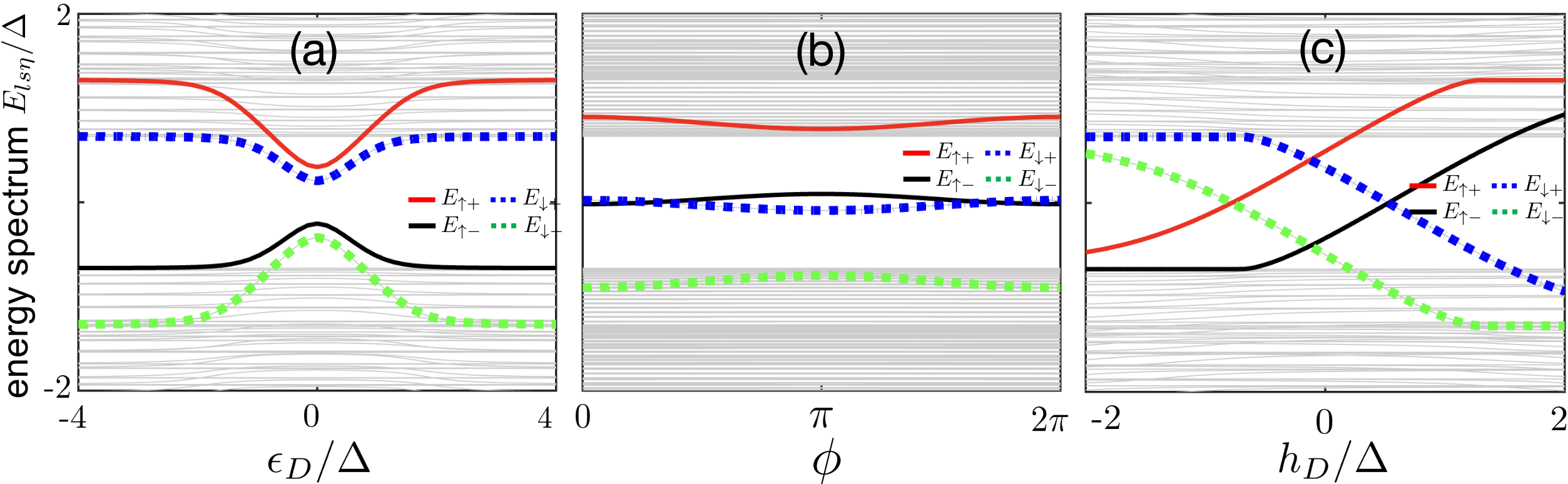}
\end{center}
\caption{(a,b,c) The  Andreev energies, $E_{ls\eta}$, numerically obtained from the diagonalization of Eq. \eqref{mvndefnvak}, are plotted as a function of (a) dot energy $\epsilon_D$ for parameter choice $(\phi/\pi,h_D/\Delta,h_L/\Delta)=(0.0,0.0,0.3)$, (b) phase difference $\phi$ with $(\epsilon_D/\Delta,h_D/\Delta,h_L/\Delta)=(0.5,0.5,0.3)$,  and (c) Zeeman energy $h_D$ with $(\phi/\pi,\epsilon_D/\Delta,h_L/\Delta)=(0.0,0.5,0.3)$, respectively. Here, the red, black, blue, and green  lines correspond to the gate-,  phase-,  and field-tunable Andreev levels $E_{\uparrow+}$, $E_{\uparrow-}$, $E_{\downarrow+}$, and  $E_{\downarrow-}$, respectively. }
\label{FIG2}
\end{figure*}

The Andreev levels \eqref{vfmk} are renormalized by  $1/[1+\tilde{\Gamma}_s(E_{s\eta})]$, making it possible to go beyond small tunnel coupling, dot energy, dot field, and Coulomb interaction limits. Figures \ref{FIG2} (a), (b), and (c) show plots of the energy levels  $E_{ls\pm}$ as a function of (a) $\epsilon_D$,  (b) $\phi$, and (c) $h_D$, respectively. Qualitatively, the Andreev levels can be understood as coming form avoided crossing due to the tunneling between the quantum dot and superconducting leads. The avoided crossing of $E=+\epsilon_D$ and $E=-\epsilon_D$ bands results in a minigap of the Andreev levels at $\epsilon_D=0$, while that of $E=\pm \epsilon_D$ and $E=sh_L+\pm \Delta$ bands repulse a bulk Bogoliubov level within the superconductor gap for $\vert \epsilon_D\vert >\Delta$. However, the avoided crossing of $E=\pm h_D$ and $E=sh_L+\pm \Delta$ bands results into a hysteresis-loop-like $h_D$ dependence [panel (c)]. Quantitatively, we try to attain some simplified expressions in several parameter regimes. 
In the  limit of large gap, i.e., $\Delta\gg h_L,h_D,\Gamma,\epsilon_D$, we have $\tilde{\Gamma}_s(E_{s\eta})\simeq 2\Gamma/\Delta\ll 1$ and the Andreev levels \eqref{vfmk} reduce to 
\begin{align} \label{gkbmkg}
    E_{s\eta}\simeq s\frac{h_D+2\Gamma h_L/\Delta}{1+2\Gamma/\Delta}+\eta \frac{\sqrt{\epsilon_D^2+4\Gamma^2\cos^2(\frac{\phi}{2})}}{1+2\Gamma/\Delta}.
\end{align}
The second term on the right-hand side of Eq. \eqref{gkbmkg} describes the presence of the minigap of the Andreev levels at $\epsilon_D=0$ due to the avoided crossing of $E=+\epsilon_D$ and $E=-\epsilon_D$ bands [Fig. \ref{FIG2} (a)]. Next, we study the limit of large dot energy ($\epsilon_D\gg \Delta,\Gamma,h_L,h_D$), large dot field ($h_D\gg \Delta,\Gamma,\epsilon_D,h_L$), and large tunneling ($\Gamma\gg \Delta,\epsilon_D,h_L,h_D$), where the Andreev levels approach the gap edges when the phase difference $\phi$ is far away from $\pm\pi$. Thus, we have $\tilde{\Gamma}(E_{s\eta})\gg 1$, and the Andreev levels  \eqref{vfmk}, in second order of $1/\tilde{\Gamma}(E_{s\eta})$, take on the form
\begin{align} \label{fvmfkk}
    E_{s\eta}-sh_L&\simeq \eta\Delta_r +(sh_D-sh_L-\eta\Delta_r)/\tilde{\Gamma}_s(E_{s\eta}) \notag \\
    &+\left[sh_L+\eta\Delta_r-sh_D+\epsilon^2_D/(2\eta\Delta_r)\right]/\tilde{\Gamma}^2_s(E_{s\eta}),
\end{align}
with $\Delta_r=\Delta \vert\cos(\frac{\phi}{2})\vert$.
For $\phi=0$, squaring Eq. \eqref{fvmfkk} leads to
\begin{widetext}
\begin{align} \label{dfvakv}
    E_{s\eta}\simeq sh_L+\eta\Delta\sqrt{1-\left[\frac{4\Gamma(sh_L-sh_D+\eta\Delta)}{4\Gamma^2+\epsilon^2_D+(sh_L-sh_D+\eta\Delta)(sh_L-sh_D+3\eta\Delta)}\right]^2}.
\end{align}
\end{widetext}
This describes the avoided crossing of $E=\pm \epsilon_D$ and $E=sh_L\pm \Delta$ bands [Fig~\ref{FIG2} (a)] and the avoided crossing of $E=\pm h_D$ and $E=sh_L\pm \Delta$ bands [Fig~\ref{FIG2}  (c)]. Noting that $\tilde{\Gamma}_s(E_{s\eta})$ becomes large when Andreev level approaches gap edges, the renormalization effect always forces the Andreev levels $E_{s+}$ and $E_{s-}$ inside $(sh_L-\Delta,sh_L+\Delta) $ even in large tunnel coupling, dot energy and dot field limits, as shown in Eq. \eqref{dfvakv}. Moreover, the renormalization effect makes it works when two "subgap" levels, $E_{\uparrow+}$ (red curve) and $E_{\downarrow-}$ (green curve) in panels (a) and (b), leak out of the superconducting gap $[-\Delta+h_L, +\Delta-h_L]$ into the continuous part
of the superconducting lead spectrum $(-h_L-\Delta,-\Delta+h_L)$ and $(+\Delta-h_L,+h_L+\Delta)$,  an effect that has no counterpart in the superconducting Anderson model~\cite{zalom2022subgap}. These "subgap" levels are  highly gate-, phase-, and field-tunable as plotted in Fig.~\ref{FIG2}, and hence play a significant role in superconducting properties, for example supercurrent  noting that a quasiparticle state of $E_{ls\eta}$ contributes to the supercurrent an amount $I_{ls\eta}=\frac{2e}{\hbar}\partial_ \phi E_{ls\eta}$.


\section{Yu-Shiba-Rusinov case} 
\label{YushibaRusinov}
In this section, we include Coulomb interaction and study the  Yu-Shiba-Rusinov case using the Hartree-Fock-Bogoliubov approximation \cite{valentini2021nontopological,lee2014spin,vecino2003josephson}.

In the presence of Coulomb interaction, the dot Hamiltonian, reads
\begin{align} 
    H_D=\sum_{s}(\epsilon_D+sh_D)n_{s}+Un_{\uparrow}n_{\downarrow}.
\end{align}
Here, $U$ is on-site Coulomb interaction and $n_s=d^{\dagger}_sd^{}_s$ is number operator of the quantum dot with spin $s$. Within the Hartree-Fock-Bogoliubov approximation, we reach
\begin{align} \label{mfmmll}
    \mathcal{H}^s_D=\epsilon_D\tau_z+sh_D+\left(\begin{array}{cc}
U\left\langle n_{-s}\right\rangle & U\left\langle sd^{}_{s} d^{}_{-s}\right\rangle \\
U\left\langle sd^{\dagger}_{-s}d^{\dagger}_{s} \right\rangle & -U\left\langle n^{}_{s}\right\rangle
\end{array}\right).
\end{align}
where $\langle \cdots\rangle$ means thermodynamic expectation. Thus, Coulomb interaction results in renormalization of dot energy, dot field, and dot pair potential  
\begin{align} \label{fdmvlfvm1}
     \epsilon^r_D=\epsilon_D+\frac{U}{2}\left(\left\langle n_{\uparrow}\right\rangle+\left\langle n_{\downarrow}\right\rangle\right),
\end{align}
\begin{align} \label{fdmvlfvm2}
     h^r_D=h_D+\frac{U}{2}\left(\left\langle n_{\downarrow}\right\rangle-\left\langle n_{\uparrow}\right\rangle\right),
\end{align}
\begin{align} \label{fdmvlfvm3}
     \Delta^r_D=-U\left\langle sd_{s} d_{-s}\right\rangle.
\end{align}
Following the same way of non-interacting case, we obtain Andreev Hamiltonian -- the low-energy  Hamiltonian of the hybrid quantum dot-nanowire-superconductor system
\begin{align} \label{ynvdknkdf}
    H_{\text{A}}=\sum_{s=\uparrow/\downarrow,\eta=+/-} E_{s\eta} A^{\dagger}_{s\eta}A^{}_{s\eta}.
\end{align}
The expressions of Andreev levels are, again, given by iterative form
\begin{align} \label{fkvldblf}
    E_{s\eta}&=s\frac{1}{1+\tilde{\Gamma}_s(E_{s\eta})}\left[h^r_D+\tilde{\Gamma}_s(E_{s\eta})h_L\right]\\
    &+\eta \frac{1}{1+\tilde{\Gamma}_s(E_{s\eta})}\sqrt{\left(\epsilon^r_D\right)^2+\left[\Delta^r_D-\Delta\cos(\phi/2)\tilde{\Gamma}_s(E_{s\eta})\right]^2}.\notag
\end{align}

To calculate the mean values in Eqs. (\ref{fdmvlfvm1}-\ref{fdmvlfvm3}), we derive the partition functions of the hybrid quantum dot and superconductor system \cite{meng2009self}. The partition function of the hybrid  system reads
\begin{align} \label{mfjkalkarfslal}
    Z^s&= \int  D[\psi_s, \bar{\psi}_s] \int D[\Psi_s, \bar{\Psi}_s]  \mathrm{e}^{-\sum_{\omega_n}S_s(i\omega_n)}.
\end{align}
The action, $S_s$ in partition function \eqref{mfjkalkarfslal} 
\begin{align} \label{fvkdmvk}
    S_s(i\omega_n)=[ \bar{\psi}_s, \bar{\Psi}_s ]\left[\begin{array}{cccc} 
    \mathcal{H}^s_D-i\omega_n  & T^{+} \\
    T^{} & \mathcal{H}^{s}_L-i\omega_n
    \end{array}\right]\begin{bmatrix}
     \psi_s \\
     \Psi_s
    \end{bmatrix},
\end{align}
with
\begin{align}
    \mathcal{H}^s_D=\epsilon^r_D\tau_z+sh_D^r+\Delta^r_D\tau_x.
\end{align}
The Grassmann variables in Nambu space read
\begin{align} \label{mskjvnavak1}
    \psi_s=\left(\begin{array}{c} d_{s} \\
    -s\bar{d}_{-s}
    \end{array}\right),\bar{\psi}_s=\left(\begin{array}{cccc} \bar{d}_{s}  & -sd_{-s} \end{array}\right),
\end{align} 
\begin{align} \label{mskjvnavak2}
    \Psi_s=\bigoplus_{jn\vec{k}}\left(\begin{array}{c}c_{ \vec{k} s} \\ -s\bar{c}_{-\vec{k} -s}  \end{array}\right),
    \bar{\Psi}_s=\bigoplus_{jn\vec{k}}\left(\begin{array}{cccc}\bar{c}_{ \vec{k} s}  &-sc_{-\vec{k} -s}  \end{array}\right).
\end{align}
In principle, we are required to diagonalize Eq. \eqref{fvkdmvk} and attain  partition function as follows
\begin{align} \label{tkmvmad}
     \ln Z^s&=\sum_{l\eta}\ln{\left(1+ e^{-\beta E_{ls\eta}}\right)}.
\end{align}
The mean values, in Eqs. (\ref{fdmvlfvm1}-\ref{fdmvlfvm3}), can be expressed by the Andreev partition function 
\begin{align}
    \left(\left\langle n_{\uparrow}\right\rangle+\left\langle n_{\downarrow}\right\rangle-1\right)=-\frac{1}{\beta}\frac{\partial}{\partial \epsilon^r_D} \ln Z^s,
\end{align}
\begin{align}
    \left(\left\langle n_{\uparrow}\right\rangle-\left\langle n_{\downarrow}\right\rangle+s\right)=-\frac{1}{\beta}\frac{\partial}{\partial h^r_D} \ln Z^s,
\end{align}
\begin{align}
    \left\langle sd_{s} d_{-s}\right\rangle=-\frac{1}{2\beta}\frac{\partial}{\partial \Delta^r_D} \ln Z^s.
\end{align}
Then,  Coulomb interaction results in self-consistent equations
\begin{align} \label{Snfdmvlfvm1e}
     \epsilon^r_D=\epsilon_D+\frac{U}{2}-\frac{U}{2\beta}\frac{\partial}{\partial \epsilon^r_D} \ln Z^s,
\end{align}
\begin{align} \label{Snfdmvlfvm2e}
     h^r_D=h_D+s\frac{U}{2}+\frac{U}{2\beta}\frac{\partial}{\partial h^r_D} \ln Z^s,
\end{align}
\begin{align} \label{Snfdmvlfvm3e}
     \Delta^r_D=-\frac{U}{2\beta}\frac{\partial}{\partial \Delta^r_D} \ln Z^s.
\end{align}
By substitution of Eq. \eqref{tkmvmad}, Eqs. (\ref{Snfdmvlfvm1e}-\ref{Snfdmvlfvm3e}) become
\begin{align} \label{rnfdmvlfvm1e}
     \epsilon^r_D=\epsilon_D+\frac{U}{2}+\frac{U}{2}\sum_{l\eta}f\left(E_{ls\eta}\right)\frac{\partial}{\partial \epsilon^r_D}E_{ls\eta},
\end{align}
\begin{align} \label{rnfdmvlfvm2e}
     h^r_D=h_D+s\frac{U}{2}-\frac{U}{2}\sum_{l\eta}f\left(E_{ls\eta}\right)\frac{\partial}{\partial h^r_D}E_{ls\eta},
\end{align}
\begin{align} \label{rnfdmvlfvm3e}
     \Delta^r_D=-\frac{U}{2}\sum_{l\eta}f\left(E_{ls\eta}\right)\frac{\partial}{\partial \Delta^r_D}E_{ls\eta}.
\end{align}
where we have used the relations
\begin{align}
    \frac{\partial}{\partial X^r_D} \ln Z^s&=-\beta\sum_{l\eta}f\left(E_{ls\eta}\right)\frac{\partial}{\partial X^r_D}E_{ls\eta}.
\end{align}
Here we can pick up anyone spin species.

Let us first switch off the quantum tunneling between quantum dot and superconductors. 
For $\Gamma=0$, we have $\Delta^r_D=0$. Thus, Eq. \eqref{fkvldblf} reduces to
\begin{align}
     E_{s\eta}=sh^r_D+\eta \sqrt{\left(\epsilon^r_D\right)^2},
\end{align}
and Eqs. \eqref{rnfdmvlfvm1e} and \eqref{rnfdmvlfvm2e} reduce to 
\begin{align} \label{ynfdmvlfvm1r}
     \epsilon^r_D=\epsilon_D+\frac{U}{2}+\frac{U}{2}\sum_{\eta}\eta f(E_{s\eta})\frac{ \epsilon^r_D}{\sqrt{\left(\epsilon^r_D\right)^2}},
\end{align}
\begin{align} \label{ynfdmvlfvm2r}
     h^r_D=h_D+s\frac{U}{2}-\frac{U}{2}\sum_{\eta}sf\left(E_{s\eta}\right).
\end{align}
We note that $E_{s+}=-E_{-s-}$ and hence we have
\begin{align}
    f\left(E_{\uparrow+}\right)&-f\left(E_{\uparrow-}\right)=f\left(-E_{\downarrow-}\right)-f\left(-E_{\downarrow+}\right)\\
    &=f\left(E_{\downarrow+}\right)-f\left(E_{\downarrow-}\right),\notag
\end{align}
\begin{align}
    1-&f\left(E_{\uparrow+}\right)-f\left(E_{\uparrow-}\right)=1-f\left(-E_{\downarrow-}\right)\\
    &-f\left(-E_{\downarrow+}\right)=-1+f\left(E_{\downarrow+}\right)+f\left(E_{\downarrow-}\right).\notag
\end{align}
Therefore, both $\epsilon^r_D$ and $h^r_D$ are independent of $s$. 
For $h_D,\epsilon_D>0$. The QPT happens at $\vert \epsilon^r_D\vert=h^r_D$. For positive $\epsilon^r_D$, we have $\epsilon^r_D=h_D^r$. By substitution of Eqs. \eqref{ynfdmvlfvm1r} and \eqref{ynfdmvlfvm2r}, we find
\begin{align}
    \epsilon^{c,+}_D=h_D-Uf(E_{\uparrow+}),
\end{align}
which at zero temperature reduces to $\epsilon^{c,+}_D=h_D$  due to $f(E_{\uparrow+})\rightarrow 0$. For negative $\epsilon^r_D$, we have $\epsilon^r_D=-h_D^r$. By substitution of Eqs. \eqref{ynfdmvlfvm1r} and \eqref{ynfdmvlfvm2r}, we find
\begin{align} 
   \epsilon^{c,-}_D
&=-h_D-U +Uf(E_{\uparrow+}) ,
\end{align}
which at zero temperature reduces to $\epsilon^{c,-}_D=-h_D-U$  due to $f(E_{\uparrow+})\rightarrow 0$.

Next, we switch on the quantum tunneling between the quantum dot and superconductors.  Figs. \ref{FIGCI} (a) and (b), respectively, plot the Andreev levels as a function of $\epsilon_D$ and $\phi$ in the presence of Coulomb interaction. We see a clear zero-energy shift (green arrows) due to the Coulomb interaction. This shift cause an enhancement of the region of the double ground state plotted in Figs. \ref{FIGCI} (c) and (d), which plot the phase diagram as a function of $\epsilon_D$ and $\phi$  for $h_L/\Delta=0$ and $h_L/\Delta=0.2$, respectively. The Coulomb interaction, preferring doublet, enhances the region of the doublet ground state [Fig. \ref{FIGCI} (c)], which can be further enhanced by the spin-split proximity effect ($h_L\neq 0$) [Fig. \ref{FIGCI} (d)]. Moreover, we find the Coulomb enhancement of the doublet ground state depends on the phase difference, $\phi$. Take $\phi=\pi$ as an example. The superconducting proximity effects from two superconductors interfere destructively. Thus, we have $\Delta^r_D=0$ and the Andreev levels \eqref{fkvldblf} reduces to 
\begin{align} \label{yifkvldblfr}
    E_{s\eta}=s\frac{1}{1+\tilde{\Gamma}_s(E_{s\eta})}\left[h^r_D+\tilde{\Gamma}_s(E_{s\eta})h_L\right]+\eta \frac{\vert \epsilon^r_D\vert }{1+\tilde{\Gamma}_s(E_{s\eta})}.
\end{align}
Read from Eq. \eqref{yifkvldblfr}, it is clear that we have
\begin{align} \label{fvdllr}
    \eta\frac{\partial}{\partial \vert \epsilon^r_D\vert}E_{s\eta}=s\frac{\partial}{\partial h^r_D}E_{s\eta}.
\end{align}
The QPT happens at zero Andreev levels.  Without loss of generality, we set $h_D,h_L>0$ and we have $E^{\uparrow-}_{\text{A}}=E^{\downarrow+}_{\text{A}}=0$ at $\vert\epsilon^r_D\vert = h^r_D+\tilde{\Gamma}(0)h_L$, where $\tilde{\Gamma}(0)=2\Gamma/\sqrt{\Delta^2-h_L^2}$. Let us consider two cases. Let us begin with $\epsilon^r_D>0$ case. By substitution of Eqs. \eqref{rnfdmvlfvm1e} and \eqref{rnfdmvlfvm2e}, we reach
\begin{align} \label{trgjorr}
   \epsilon^{c,+}_D
=\tilde{\Gamma}(0)h_L+h_D-\frac{U}{2}\sum_{l\eta}f(E_{l\uparrow\eta})\left(\frac{\partial}{\partial h^r_D}+\frac{\partial}{\partial \epsilon^r_D}\right)E_{l\uparrow\eta}.
\end{align}
 Next, we study $\epsilon^r_D<0$ case.
By substitution of Eqs. \eqref{rnfdmvlfvm1e} and \eqref{rnfdmvlfvm2e}, we reach
\begin{align} \label{ytrgjorr}
   \epsilon^{c,-}_D
&=-h_D-\tilde{\Gamma}(0)h_L-U \\
&+\frac{U}{2}\sum_{l\eta}f(E_{l\uparrow\eta})\left(\frac{\partial}{\partial h^r_D}-\frac{\partial}{\partial \epsilon^r_D}\right)E_{l\uparrow\eta}.\notag 
\end{align}

For simplicity, we consider weak tunnel coupling, small dot energy, and small Coulomb interaction limits, where the dot energy and field dependence of quasiparticle energy $E_{ls\eta}$ is dominated by Andreev levels, i.e.,
\begin{align}
    \frac{\partial}{\partial h^r_D}E_{l\uparrow\eta}\simeq \delta_{l,1}\frac{\partial}{\partial h^r_D}E_{\uparrow\eta},
\end{align}
\begin{align}
    \frac{\partial}{\partial \epsilon^r_D}E_{l\uparrow\eta}\simeq \delta_{l,1}\frac{\partial}{\partial \epsilon^r_D}E_{\uparrow\eta}.
\end{align}
By substitution of Eq. \eqref{fvdllr}, Eq. \eqref{trgjorr} becomes
\begin{align} \label{trgjtorr}
   \epsilon^{c,+}_D
=\tilde{\Gamma}(0)h_L+h_D-Uf(E_{\uparrow+})\frac{\partial}{\partial h^r_D}E_{\uparrow+},
\end{align}
which at zero temperature reduces to $\epsilon^{c,+}_D=h_D+\tilde{\Gamma}(0)h_L$ due to $f(E_{\uparrow+})\rightarrow 0$. Therefore, positive critical dot energy \eqref{trgjtorr}, at $\phi=\pi$ and $T=0$K, is independent of $U$  but relies on $h_D$, $h_L$, and $\Gamma$ [red circles of Figs. \ref{FIGCI} (c) and (d)]. Moreover, by substitution of Eq. \eqref{fvdllr}, Eq. \eqref{ytrgjorr} becomes
\begin{align} \label{ytrgjtorr}
   \epsilon^{c,-}_D
&=-h_D-\tilde{\Gamma}(0)h_L-U +Uf(E_{\uparrow+})\frac{\partial}{\partial h^r_D}E_{\uparrow+},
\end{align}
which at zero temperature reduces to $\epsilon^{c,-}_D=-h_D-\tilde{\Gamma}(0)h_L-U$ due to, again, $f(E_{\uparrow+})\rightarrow 0$. The negative critical dot energy \eqref{ytrgjtorr} is shifted by Coulomb interaction [black circles of Figs. \ref{FIGCI} (c) and (d)].

%

\end{document}